\documentclass[10pt,a4paper,twoside]{article}
\usepackage{epsfig}
\usepackage{baltlat5}
\usepackage{wrapfig}
\newcommand{\gtrsim}{\mathrel{\hbox{\rlap{\hbox{\lower4pt\hbox{$\sim$}}}\hbox{$>$}}}}
\newcommand{\lesssim}{\mathrel{\hbox{\rlap{\hbox{\lower4pt\hbox{$\sim$}}}\hbox{$<$}}}}

\begin{document}
\vspace{0.5mm}

\setcounter{page}{1}
\vspace{5mm}




\titlehead{Baltic Astronomy, vol.\ts 14, XXX--XXX, 2005.}

\titleb{NEW CANDIDATE EHB STARS IN THE OPEN CLUSTER 
NGC~6791: LOOKING LOCALLY INTO THE UV-UPTURN PHENOMENON}


\begin{authorl}
\authorb{L.M.~Buson}{1}
\authorb{E.~Bertone}{2}
\authorb{A.~Buzzoni}{3}, and
\authorb{G.~Carraro}{4,5,6}
\end{authorl}

\begin{addressl}
\addressb{1}{INAF - Osservatorio di Padova, Italy}
\addressb{2}{INAOE, Puebla, Mexico}
\addressb{3}{INAF - Osservatorio di Bologna, Italy}
\addressb{4}{Dept. de Astron. Univ. de Chile, Santiago. Chile} 
\addressb{5}{Astronomy Dept., Yale Univ., USA}
\addressb{6}{Dip. di Astronomia, Univ. di Padova, Italy}
\end{addressl}

\submitb{Received 2005 August 1}

\begin{abstract}
Relying on U,B imagery at the Italian Telescopio Nazionale Galileo (TNG), 
we report here the discovery of a sample of 13 new UV-bright post-HB candidate 
stars in the field of the galactic open cluster NGC~6791. Owing to its 
super-solar metal content ([Fe/H]~$\gtrsim 0.2$~dex) and estimated 
age ($t \gtrsim 8$~Gyr), this cluster represents the nearest and ideal stellar 
aggregate to match the distinctive properties of the evolved stellar populations
possibly ruling the UV-upturn phenomenon in elliptical galaxies
and bulges of spirals. 

Our ongoing spectroscopic follow-up of this unique UV-bright sample 
will allow us to assess -- once cluster membership of the candidates is 
properly checked-- the real nature (e.g. SdB, SdO, AGB-manqu\'e or EHB stars) 
of these hot sources, and their link with the ultraviolet excess emerging 
from low-mass, metal-rich evolutionary environments of external galaxies.
\end{abstract}


\resthead{\LaTeX\ style for Baltic Astronomy}{L.M.~Buson et al.}

\sectionb{1}{INTRODUCTION}

Since its  early discovery (Code 1969), the so-called UV-upturn 
phenomenon in old stellar populations of ellipticals and spiral bulges 
(namely the abrupt rise in the UV continuum emission shortward of 
$\lambda\sim 2,000$~\AA) has been the subject of growing theoretical
analyses intended to establish its origin and evolution (see e.g.\ 
Greggio \& Renzini 1990 and O'Connell 1999, for a review).

Both theory and observations currently seem to converge 
towards the ``Extreme Horizontal Branch'' (EHB) scenario as the main
responsible for the phenomenon (Dorman et al.\ 1995; Brown 2004).
If this is the case, models show that hot HB stars with Helium core mass 
$M_{\rm core} \lesssim 0.52~M_\odot$ can escape the standard Post-HB evolution
(that would culminate with the planetary-nebula event at the end of the 
asymptotic giant branch evolution), and directly reach the high-temperature 
region of the H-R diagram ($T_{\rm eff} \gtrsim 30\,000$~K) to fade 
then along the white-dwarf cooling sequence (Dorman et al.\ 1993).

In this framework, the role of metallicity cannot yet be
confidently assessed, however, as we face two conflicting scenarios 
relying either on a metal-poor evolution (naturally giving rise 
to a blue HB morphology, see Park \& Lee 1997) or 
a metal-rich case, where the onset of UV emission needs at least a 
fraction of stars to exceed some critical threshold in [Fe/H]
(Greggio \& Renzini 1990; Bressan et al.\ 1994; Buzzoni 1995; 
Dorman et al.\ 1995). 
In this regard, one should be aware that even the high-resolution 
UV spectroscopy provided by FUSE for the UV-brightest elliptical NGC~1399
turned out to be inadequate to solve the problem, as it mainly probes 
the photospheric abundance of hot stars, likely perturbed by diffusion 
effects and therefore not fully indicative of the true metallicity of the 
whole galaxy stellar population (Brown et al.\ 2002).

\sectionb{2}{NGC 6791: THE UNEXPECTED SHORTCUT}

Photometric observations of evolved UV-bright stars in external galaxies 
are still confined to the relevant case of M31 and its satellite system 
(e.g.\ Bertola et al.\ 1995; Brown et al.\ 1998, 2000), and no suitable 
spectroscopy for single stars is available to date.

Surprisingly enough, the closeby Galactic open cluster NGC~6791, 
less than 5~kpc away (Friel 1995; Carraro et al.\ 1999), turned out to be 
a highly valuable candidate to address the issue of the EHB UV-bright stars, 
standing out as a sort of backyard ``Rosetta Stone'' to assess the 
UV emission of spheroids much farther away. 
This cluster is actually one of the brightest ($L_V \sim 6.3\,10^3~L_\odot$), 
oldest ($t \gtrsim 8$~Gyr) and 
metal-rich ([Fe/H]$\gtrsim +0.2$) ones (King et al.\ 2005; Stetson et al.\ 2003; 
Carraro et al.\ 1999), and hosts a significant fraction of sdB/O stars
(Kaluzny \& Rucinski 1995; Kaluzny \& Udalski 1992, see Fig.~1) interpreted by 
Yong et al.\ (2000) as EHB stars with $T_{\rm eff}$ in the range 24--32\,000~K, 
as confirmed by ground and space-borne (UIT and HST) observations 
(Liebert et al.\ 1994; Landsman et al.\ 1998).

\sectionb{3}{OBSERVATIONS AND REDUCTIONS}

In 2003 we started a specific observing programme aiming at imaging
a large field of NGC~6791 in the Johnson U,B wavebands with the 
LRS FOSC camera at the 3.5~m Italian Telescopio Nazionale Galileo
(TNG) at La Palma. Taking advantage of its 0.275~arcsec/px scale and its wide 
($9.4 \times 9.4$~arcmin) field of view, this imager allowed us to carry out a
suitable ($16 \times 16$ arcmin) and accurate ($\pm 0.01$~mag) survey of the
cluster region. Resulting CMDs, based on 330~sec B and 1200~sec U exposures,
are shown in Fig.~2. Data have been reduced with IRAF packages CCDRED, 
DAOPHOT, ALLSTAR and PHOTCAL, making use of the point spread 
function method (Stetson 1987).

\vbox{
\begin{minipage}{0.43\hsize}
\psfig{figure=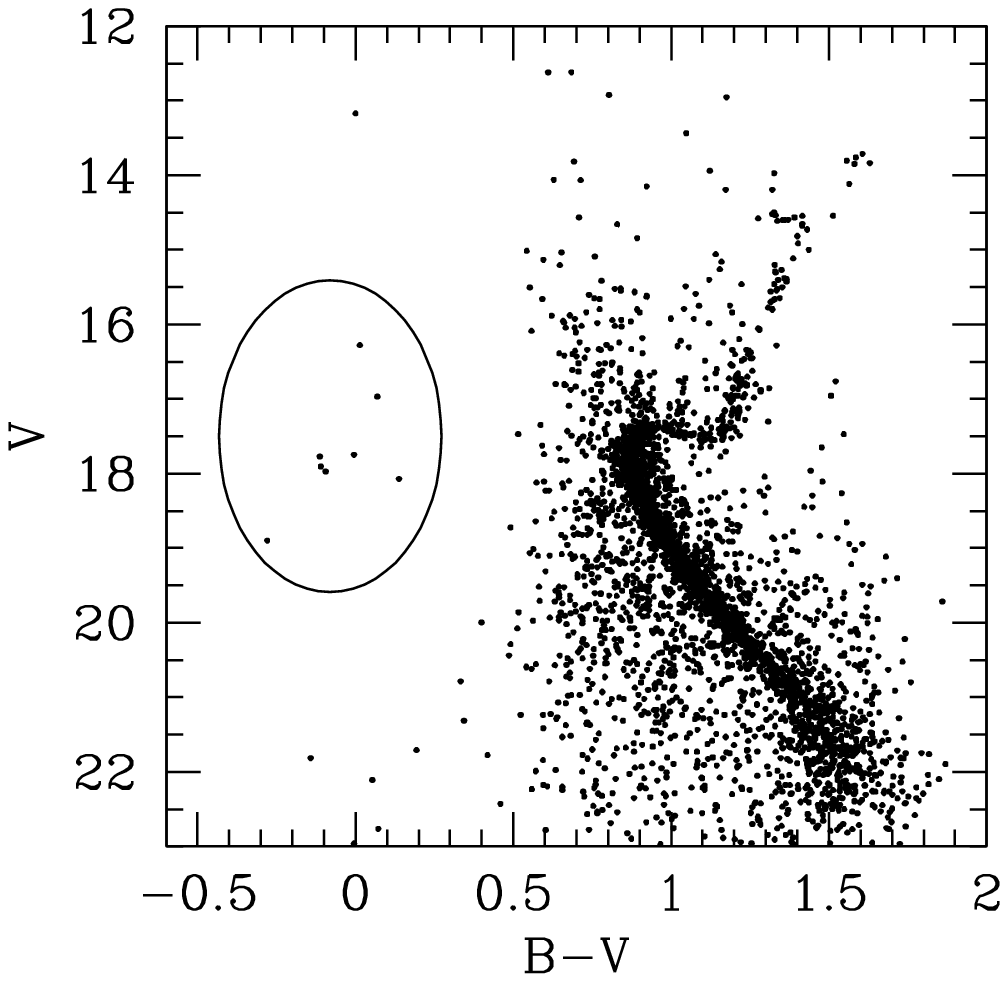,width=\hsize,clip=}
\end{minipage}
\hfill
\begin{minipage}{0.48\hsize}
\captionb{1}
{The Kaluzny \& Rucinski (1995) c-m diagram of NGC~6791, based on CCD photometry at the
ESO 2.1m telescope (only best photometric sample plotted here). Yong et al.\ (2000) 
provided a good fit to these data with a ([Fe/H], t) = (0.33~dex, 8~Gyr) model,
predicting however a red HB morphology clumped about $(B-V) \simeq 1.3$, and
clearly missing the eight hot (supposedly EHB) stars singled out in the plot.}
\end{minipage}
}

\vbox{
\begin{minipage}{0.56\hsize}
\psfig{file=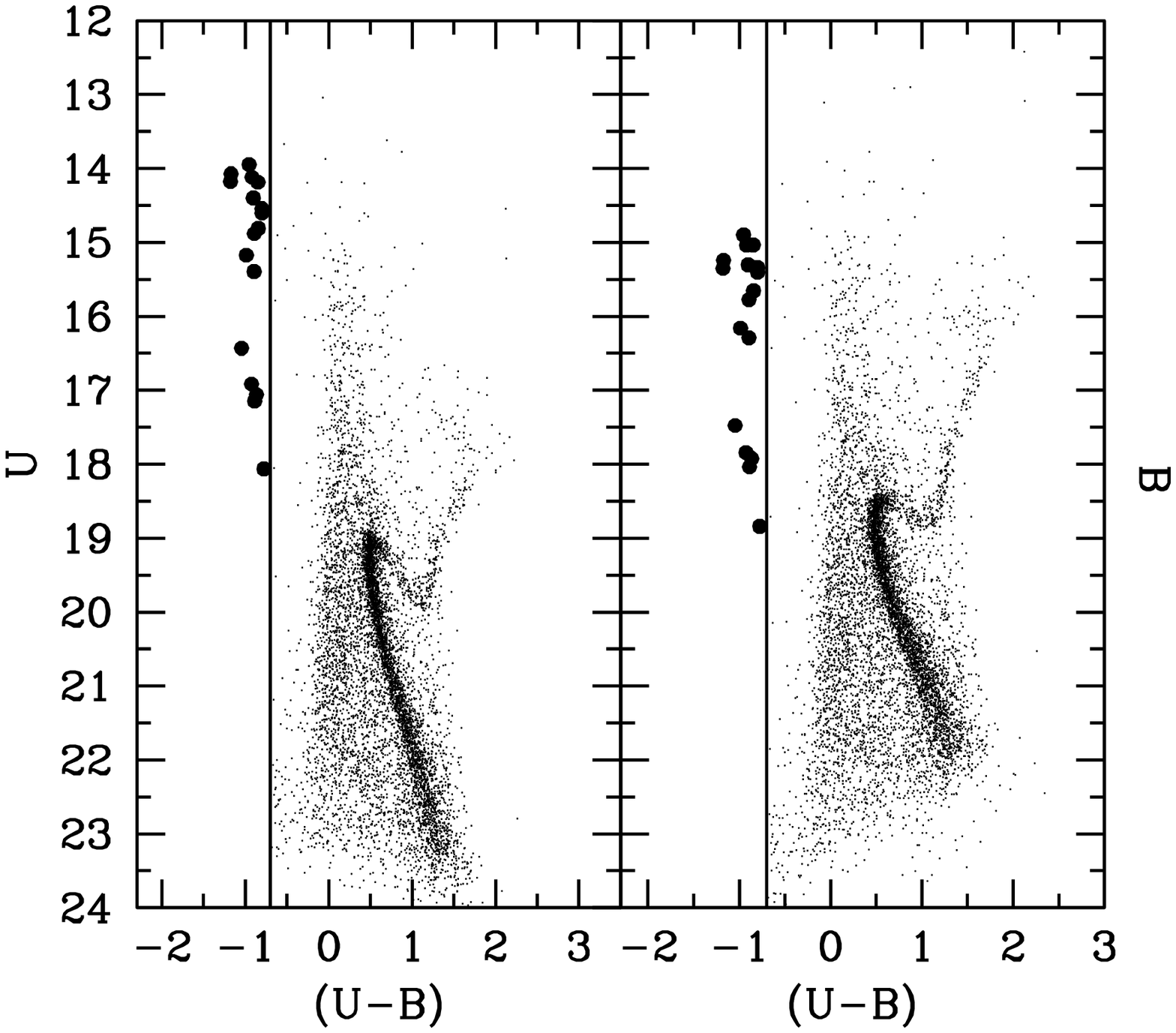,width=\hsize,clip=}
\end{minipage}
\hfill
\begin{minipage}{0.35\hsize}
\captionb{2}{Combined U vs.\ U--B and B vs.\ U--B diagrams of the field of
NGC~6791 explored with TNG LRS. Big solid dots mark our 13 newly
discovered UV-bright sources with (U--B)~$<-0.7$ (vertical line) together with four
objects (the clump about U$\sim 17$) including three member sdB and one
sdO from the sample of faint blue stars by Kaluzny \& Udalski (1992).}
\end{minipage}
}

\sectionb{4}{TOWARDS THE FUTURE}

The analysis of Fig.~2 shows a cleanly detected sample of 13 new UV-bright
objects, bluer than (U--B)~$< -0.7$, and consistent with sdB/O stars, that sums up to 
the four EHB candidates (the clump of stars around $U \sim 17$) previously found 
in this color range by Kaluzny \& Udalski (1992).\footnote{Although about 2~mag brighter
than the Kaluzny \& Udalski (1992) candidates, our targets might still be consistent
with an sdB/O classification, recalling the large spread in the bolometric correction
to $U$ and $B$ magnitudes and the allowed temperature range compatible with the observed 
scatter in the (U--B) color (cf.\ e.g.\ Johnson 1966).}

Our current observational campaign, making use of several spectroscopic facilities
at TNG and other telescopes, will provide us with both low- ($\sim 10$~\AA\ FWHM) 
and mid- ($\sim 6$~\AA) resolution spectra along the full optical range 
($\lambda\lambda$ 3500---8000~\AA) for these newly identified targets, assessing
their cluster membership and, for those positive cases, allowing us to settle  
the effective temperature and surface gravity of each star by fitting the observed 
spectral energy distribution with the new UVBLUE and BLUERED synthetic spectral 
libraries (Rodriguez-Merino et al.\ 2005; Bertone et al.\ 2003) of appropriate 
metallicity. Among others, such a complete sample of {\it bona fide} EHB stars will 
also provide a first reliable estimate of the lifetime and global UV energetic 
budget associated to the EHB evolution relying on the so-called ``Fuel Consumption 
Theorem'' of Renzini \& Buzzoni (1986).   

\References

\refb
Bertola~F., Bressan~A., Burstein~D., Buson~L.~M., Chiosi~C.,
di Serego Alighieri~S. 1995, ApJ, 438 680

\refb
Bertone~E., Buzzoni~A., Rodriguez-Merino~L.~H., Chavez~M. 2003, in
{\em Modelling of Stellar Atmospheres}, IAU Symp. 210, eds. N.~E.~Piskunov, 
W.~W.~Weiss \& D.~F.~Gray (ASP: San Francisco)

\refb
Bressan~A., Chiosi, C., Fagotto, F. 1994, ApJS, 94, 63 

\refb
Brown~T.~M. 2004, ApSS, 291 215

\refb
Brown~T.~M., Ferguson~H.~C., Deharveng~J.-M., Jedrzejewski~R.~I. 1998, ApJ,
508, L139 

\refb
Brown~T.~M., Bowers~C.~W., Kimble~R.~A., Sweigart,~A.~V., Ferguson~H.~C. 2000, 
ApJ, 532, 308 

\refb
Brown~T.~M., Ferguson~H.~C., O'Connell~R.~W., Ohl~R.~G. 2002, ApJ, 568 L19

\refb
Buzzoni~A. 1995, ApJS, 98, 69


\refb
Carraro~G., Girardi~L., Chiosi~C. 1999, MNRAS, 309, 430

\refb
Code~A.~D. 1969, PASP, 81, 475 

\refb
Dorman~B., Rood~R.~T., O'Connell~R.~W. 1993, ApJ, 419, 596

\refb
Dorman~B., O'Connell~R.~W., Rood~R.~T. 1995, ApJ, 442, 105 

\refb
Friel~E.~D. 1995, ARAA, 33, 381

\refb
Greggio~L., Renzini~A. 1990, ApJ, 364, 35 

\refb
Johnson~H.L. 1966, ARA\&A, 4, 193
\refb
Kaluzny~J., Rucinski~S.~M. 1995, A\&AS, 114, 1
 
\refb
Kaluzny~J., Udalski~A. 1992, AcA, 42, 29 

\refb
King~I.~R., Bedin~L.~R., Piotto~G., Cassisi~S., Anderson~J. 2005, AJ, 130, 626

\refb
Landsman~W. Bohlin~R.~C., Neff~S.~G., O'Connell~R.~W., Roberts~M.~S.,
Smith~A.~M., Stecher~T.~P. 1998, AJ, 116, 789

\refb
Liebert~J., Saffer~R.~A., Green~E.~M. 1994, ApJ, 107, 1408

\refb
O'Connell~R.~W. 1999, ARAA, 37, 603

\refb
Park~J.-H., Lee~Y.-W. 1997, ApJ, 476, 28

\refb
Renzini~A., Buzzoni~A. 1986 in {\em Spectral Evolution of Galaxies}, 
eds. C.~Chiosi \& A.~Renzini (Dordrecht: Reidel), p. 195

\refb
Rodriguez-Merino~L.~H., Chavez~M., Bertone~E., Buzzoni~A. 2005, ApJ, 626, 411 

\refb
Stetson~P.~B. 1987, PASP, 99, 191

\refb
Stetson~P.~B., Bruntt~H., Grundahl~F. 2003, PASP, 115, 413

\refb
Yong~H., Demarque~P., Yi~S. 2000, ApJ, 539 928

\end{document}